# Study of Structural Properties in Complex Fluids by Addition of Surfactants Using DPD Simulation

Estela Mayoral, Eduardo Nahmad-Achar, José Manuel Martínez-Magadán, Alejandro Ortega and Ismael Soto

**Abstract** In this work we study the tertiary structure of ionic and surfactant when the pH in the system is modified using electrostatic dissipative particle dynamics simulations (DPD). The dependence with pH and kind of surfactant is presented. Our simulations reproduce the experimental behavior reported in the literature. The scaling for the radius of gyration with the size of the molecule as a function of pH is also obtained.

## 1 Introduction

Modification and control of structural and interfacial properties between the different components in confined mixed systems e.g. rock/water/oil, by the use of chemical additives, is nowadays an important research area in order to enhanced oil recovery (EOR) retained in the porous rock (Abdel-Wali 1996). EOR is based on the fact that the incorporation of other components into this complex system modifies the collective properties amongst them. The performance of the additives in the system is in strong correlation with their tertiary structure, the characteristics

E. Mayoral (✉)
Instituto Nacional de Investigaciones Nucleares (ININ), Carretera México-Toluca (S/N),
CP 52750 La Marquesa Ocoyoacac, Estado de México, Mexico
e-mail: estela.mayoral@inin.gob.mx

J. M. Martínez-Magadán · A. Ortega · I. Soto
Instituto de Ciencias Nucleares, Universidad Nacional Autónoma de México (UNAM),
Apartado Postal 70-543, 04510 México, D.F., Mexico

E. Nahmad-Achar
Instituto Mexicano del Petróleo (IMP), Eje Central Lázaro Cárdenas S/N,
México, D.F., Mexico

in the media, and the properties of oil/aqueous interfaces (Hansson and Lindman 1996). In particular, ionic strength, pH, temperature, and pressure play an important role in the behavior of these systems. Understanding how these conditions in the media modify the tertiary structure of the additives is then a fundamental task in order to design new surfactants ad hoc and to develop optimal formulations. The experimental study of the conformation of macromolecules is usually done by dynamic light scattering (DLS), but in many occasions this is complicated due to the different sizes of the molecules involved in multicomponent systems (González-Melchor et al. 2006). Alternatively, numerical simulation offers a viable option to help in the design of new additives that could give a good performance in specific situations, even under extreme conditions that are impossible to handle in laboratory. In these kinds of complex systems, where many particles with different sizes undergo interactions at different time scales, the mesoscopic simulation has prevented to be a good alternative (Español and Warren 1995; Groot and Warren 1997). One of the most important mesoscopic simulation approaches is the dissipative particle dynamics (DPD), introduced by Hoogerbruge and Koelman (1992). In this work we present a study of the radius of gyration for ionic surfactant in different pH conditions. We use electrostatic dissipative particle dynamics simulations to study its dependence with the pH and the kind of polyelectrolyte added. Our simulations reproduce the experimental information reported in the literature (Griffiths et al. 2004). The correspondence between the partial charge on the electrolyte ($\theta$) and the pH is established. Additionally, study the scaling of the radius of gyration with the size of the polymer ($N$) at different pH conditions. The scaling exponent $v$ at different pH is presented.

## 2 Methodology

Dissipative Particle Dynamics (DPD) is a *coarse graining* approach which consists of representing a complex molecule (polymer or surfactant) by soft spherical beads joined by springs, interacting through a simple pair-wise potential and thermally equilibrated through hydrodynamics (Español 2002; Español and Warren 1995). The beads follow Newton's equations of motion $\frac{dr_i}{dt} = v_i; \frac{dv_i}{dt} = f_i$, where the force $f_i$ on a bead is constituted by three pair-wise additive components $fi = \sum_j (f_{ij}^C + f_{ij}^D + f_{ij}^R)$. The sum runs over all nearby particles within a distance $R_c$. The conservative force is defined as $f_{ij}^c = a_{ij}\omega^c(r_{ij})\hat{r}_{ij}$ for $r_{ij} < r_c$ and zero elsewhere. $a_{ij}$ is a parameter which measures the maximum repulsion between particles $i$ and $j$, $r_{ij} = r_i - r_j$ and $r_{ij} = r_{ij}|\hat{r}_{ij}|$; the weight function $\omega_c(r_{ij})$ is given by $\omega^c(r_{ij}) = 1 - \frac{r_{ij}}{r_c}$ for $r_{ij} < r_c$, and zero elsewhere. This conservative repulsion force derives from a soft interaction potential and there is no hard-core divergence of this force as in the case of the Lennard-Jones potentials which allows for a more efficient scheme of integration. When we need to introduce a more complex molecule, such as a polymer, we use beads joined by springs, so we also have an extra spring force given by $f_{ij}^s = -Kr_{ij}$ if $i$ is connected with $j$. The dissipative and random standard DPD forces are given Eq. (1):

$$\boldsymbol{f}_{ij}^D = -\gamma \omega^D(r_{ij})(\hat{\boldsymbol{r}}_{ij} \cdot \boldsymbol{v}_{ij})\hat{\boldsymbol{r}}_{ij} \text{ and } \boldsymbol{f}_{ij}^R = \sigma \omega^R(r_{ij})(\theta_{ij} 1/\sqrt{\delta_t})\hat{\boldsymbol{r}}_{ij} \qquad (1)$$

where $\delta_t$ is the integration time step, $\boldsymbol{v}_{ij} = \boldsymbol{v}_i - \boldsymbol{v}_j$ is the relative velocity, $\sigma$ is the amplitude of noise and $\theta_{ij}$ is a random Gaussian number with zero mean and unit variance. $\gamma$ and $\sigma$ are the dissipation and the noise strengths respectively, while $\omega^D(r_{ij})$ and $\omega^R(r_{ij})$ are dimensionless weight functions. These quantities are not independent: they are related by the fluctuation–dissipation theorem (Español and Warren 1995) by $\gamma = \frac{\sigma^2}{2k_B T}$, and therefore $\omega^D(r_{ij}) = \left[\omega^R(r_{ij})\right]^2$. This relationship maintains the temperature constant and preserves the total energy in the system. Here $k_B$ is the Boltzmann constant and T is the temperature. The consistency of the results depends principally on how well the essential characteristics for a group of atoms are included into the interaction parameters $a_{ij}$ between the DPD beads. Groot and Warren (1997) established a simple functional form of the conservative repulsion in DPD ($a_{ij}$) in terms of the Flory–Huggins $\chi$-parameter theory. Since then, the use of solubility parameters for the calculation of $\chi$-parameters in order to get the repulsive parameters for DPD simulations is commonly used. The electrostatic interactions were introduced into DPD via two different methods, by Groot (2003) and by González-Melchor et al. (2006). In both cases the point charge at the center of the DPD particle is replaced by a charge distribution along the particle. Groot (2003) solves the problem by calculating the electrostatic field on a grid. González-Melchor et al. (2006) solve the problem adapting the standard Ewald method to DPD particles.

## 3 Simulation Details

All simulations were carried out using a DPD electrostatic code. The electrostatic interactions were calculated following González-Melchor et al. (2006). We used dimensionless units denoted with an asterisk; the masses were all equal to 1. According with Groot and Warren (1997), the values for $\sigma$ and $\gamma$ were established as 3 and 4.5 respectively, with this $k_B T^* = 1$. The total average density in the system was $\rho^* = 3.0$. The simulation box was cubic with $Lx = Ly = Lz = 8.5$. Periodic boundary conditions were imposed in all directions. The time step used was set to $\Delta t = 0.04$ and we performed 25 blocks of simulations with 10,000 steps each one, and the properties were calculated by averaging over the last 10 blocks. The structure for the molecules of ionic dispersants called was mapped as shown in Figs. 1 and 2. The repulsive interaction parameters were obtained through the solubility parameters for the individual monomers indicated in Fig. 1 and using the Flory–Huggins parameters $\chi$. The parameters $a_{ij}$ were: for equal DPD beads $a_{ii} = 78$, and for different beads $a_{AB} = 78.22$, $a_{BC} = 78.55$, $a_{CA} = 78.25$, $a_{AW} = 182.33$, $a_{BW} = 169.04$, $a_{CW} = 250.13$, where the subscripts correspond to the DPD beads shown in Fig. 1, and the subscript $W$ corresponds to the interaction with water. Monomers in a polymer were joined by Hookean springs with spring constant $K = 100$ and with equilibrium distance $r_{eq} = 0.7$.

**Fig. 1** Structure and mapping for PAA

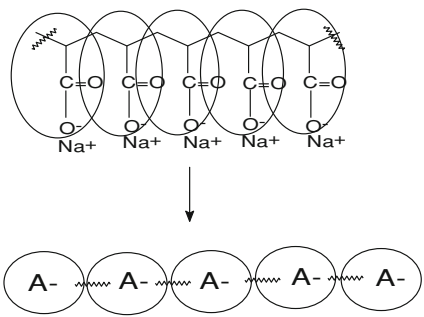

## 4 Results and Discussion

We studied the radius of gyration of an ionic surfactant (PAA) at different values of pH. The structure for this poly-electrolyte is shown in Fig. 1. The PAA, is a polyacrylic dispersant constituted by $N$ monomeric carboxylic units. This is a weak poly-acid with $pK_a$ 5.47. The carboxylic monomeric units could be neutral (if they are protonated) or negatively charged (if they are deprotonated). The degree of ionization depends on the pH of the medium: when we modify the pH some of the monomers stay uncharged ($N_0$) and the rest become negatively charged ($N^- = N - N_0$). We can express the relationship between the pH and the charge fraction $\theta$ as: $pH = \log(\theta/(1-\theta)) + pK_a$ where $\theta$ is the ratio between the number of deprotonated ($N^-$) carboxylic monomeric units and the total number of monomeric units ($N$). We performed simulations for PAA molecules of different sizes ($N = 12$, 24, 48), and for different values of pH (1, 4.62, 5.25, 5.47, 5.79, 5.94, 6.31, 14); the results for the radius of gyration as a function of pH are shown in Fig. 2.

We can observe that the radius of gyration ($Rg$) increase with pH, as expected, due to the repulsive intramolecular charge. Setting $Rg = aN^\nu$ with $\nu$ the scaling exponent and $a$ the Flory ratio, we obtain the values shown in Table 1 for different charge fractions $\theta$.

Theoretical analysis has shown that, in a good solvent, $\nu = 3/5 = 0.6$ (De Gennes 1979); looking at Table 1 and Fig. 2 we see that the scaling exponent obtained for $\theta = 0.375$ is $\nu = 0.593$. This is the closest value obtained compared with the theoretical value corresponding with the greatest value for $a = 0.556$, indicating that the molecule is completely extended as is expected for a good solvent.

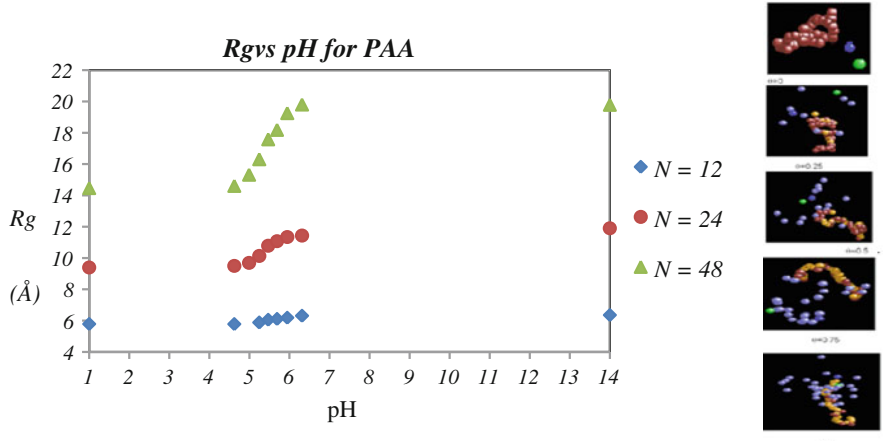

**Fig. 2** Rg for PAA as a function of pH and equilibrium configurations at different $\theta$

**Table 1** Scaling exponent $v$ for PAA at different $\theta$

| $\theta$ | 0 | 0.125 | 0.375 | 0.5 | 0.675 | 0.75 | 0.875 | 1 |
|---|---|---|---|---|---|---|---|---|
| v | 0.643 | 0.631 | 0.593 | 0.740 | 0.745 | 0.795 | 0.802 | 0.779 |
| a | 0.449 | 0.469 | 0.556 | 0.403 | 0.404 | 0.352 | 0.436 | 0.407 |

**Table 2** Simulation and experimental results for the Rg of PAA  *NA* not available

| PM (gr/mol) | 900 | 1800 | 3600 | 20,000 | 770,000 |
|---|---|---|---|---|---|
| N | 12 | 24 | 48 | 267 | 10 267 |
| Rg simulation (DPD Units) | 0.895 | 1.454 | 2.239 | 6.970 | 77.550 |
| Rg simulation (Å) | 5.7838 | 9.394 | 14.464 | 45.05 | 500.979 |
| Rg experimental (Å) | NA | NA | NA | 50 | 513 |

Table 2 shows the estimated results of the radius of gyration, both in DPD units and in Å, for different molecular weights using the scaling approach obtained with the DPD simulations. As well as the experimental data reported. The simulation with $N = 48$ corresponds with a $MW = 3600$ Da. We observed a very good agreement with the experimental results.

## 5 Conclusions

Electrostatic DPD simulations prove to be very useful in the study of the tertiary structure of polyelectrolyte, by studying the radius of gyration as a function of pH for various molecular lengths. The behavior of the ionic surfactant PAA has been obtained in very good agreement with experimental results. The correct scaling exponent $\nu$ is also obtained for the ionic surfactant. The quantitative prediction for the radius of gyration using DPD electrostatic simulations and scaling arguments corresponds with the experimental data obtained by DLS experiments.